%% file: main.tex
\documentclass[conference]{IEEEtran}
\IEEEoverridecommandlockouts
\usepackage{cite}
\usepackage{amsmath,amssymb,amsfonts,algorithm,algpseudocode}
\usepackage{graphicx}
\usepackage{textcomp}
\usepackage{xcolor}
\usepackage[backref=page]{hyperref}
\usepackage{cleveref}
\def\BibTeX{{\rm B\kern-.05em{\sc i\kern-.025em b}\kern-.08em
    T\kern-.1667em\lower.7ex\hbox{E}\kern-.125emX}}
\begin{document}

\title{Physically Interpretable Representation and Controlled Generation for Turbulence Data\\
\thanks{This material is based upon work supported by the Department of Energy’s National Nuclear Security Administration under
Award Number DE-NA0003968.}
}

\author{\IEEEauthorblockN{1\textsuperscript{st} Tiffany Fan}
\IEEEauthorblockA{\textit{Stanford University} 
}
\and
\IEEEauthorblockN{2\textsuperscript{nd} Murray Cutforth}
\IEEEauthorblockA{\textit{Stanford University} 
}
\and
\IEEEauthorblockN{3\textsuperscript{rd} Marta D'Elia}
\IEEEauthorblockA{\textit{Atomic Machines and Stanford University
} 
}
\and
\IEEEauthorblockN{4\textsuperscript{th} Alexandre Cortiella}
\IEEEauthorblockA{\textit{National Renewable Energy Laboratory} 
}
\and
\IEEEauthorblockN{5\textsuperscript{th} Alireza Doostan}
\IEEEauthorblockA{\textit{University of Colorado Boulder} 
}
\and
\IEEEauthorblockN{6\textsuperscript{th} Eric Darve}
\IEEEauthorblockA{\textit{Stanford University} 
}
}
\maketitle
\IEEEpeerreviewmaketitle


\input{./GMVAE-dimension-reduction/sections/1-introduction.tex}

\input{./GMVAE-dimension-reduction/sections/2-statement.tex}
\input{./GMVAE-dimension-reduction/sections/3-method.tex}
\input{./GMVAE-dimension-reduction/sections/5-experiments.tex}

\input{./GMVAE-dimension-reduction/sections/6-conclusion.tex}
\bibliographystyle{IEEEtran}
\bibliography{main}

\end{document}

%% file: GMVAE-dimension-reduction/sections/1-introduction.tex
\section{Introduction}
\label{sec:introduction}

Computational Fluid Dynamics (CFD) is central to fluid mechanics, offering precise simulations of fluid behavior through partial differential equations (PDEs). Traditional CFD methods, such as those based on finite difference and finite volume schemes, are resource-consuming, especially for high-fidelity simulations of complex flows. Understanding such datasets presents unique challenges due to their high dimensionality, inherent stochasticity, and limited data availability.

To address these challenges, this work explores data-driven approaches to encode high-dimensional scientific data into low-dimensional, physically meaningful representations. By leveraging these representations, we aim to uncover patterns, enable clustering, and facilitate generative modeling for complex flows.

%% file: GMVAE-dimension-reduction/sections/2-statement.tex


This work extends a Gaussian Mixture Variational Autoencoder (GMVAE) \cite{jiang2016variational} to:
\begin{itemize}
    \item extract meaningful latent representations while preserving physical structures;
    \item cluster experiments based on similar physical states, such as pressure and temperature; and
    \item facilitate generative modeling to provide low-fidelity datasets and insights into engineering systems.
\end{itemize}

In addition, a robust and interpretable framework for analyzing complex engineering systems should preserve underlying physical characteristics in the data. We introduce a novel quantitative metric for physical interpretability. This metric evaluates the smoothness of physical quantities, such as pressure, across the latent manifold by analyzing their projection onto the eigenvectors of a graph Laplacian. Combined with the GMVAE, this framework offers a robust solution for understanding and analyzing complex scientific datasets.

%% file: GMVAE-dimension-reduction/sections/3-method.tex
\section{Methodology}
\label{sec:method}

\subsection{Model Assumptions}
\label{subsec:model_assumptions}

We leverage and extend a GMVAE for dimensionality reduction and clustering, assuming that high-dimensional data reside on a low-dimensional manifold that can be clustered based on physical properties or experimental conditions. \begin{figure}[tbh]
    \centering
    \includegraphics[width=0.9\linewidth]{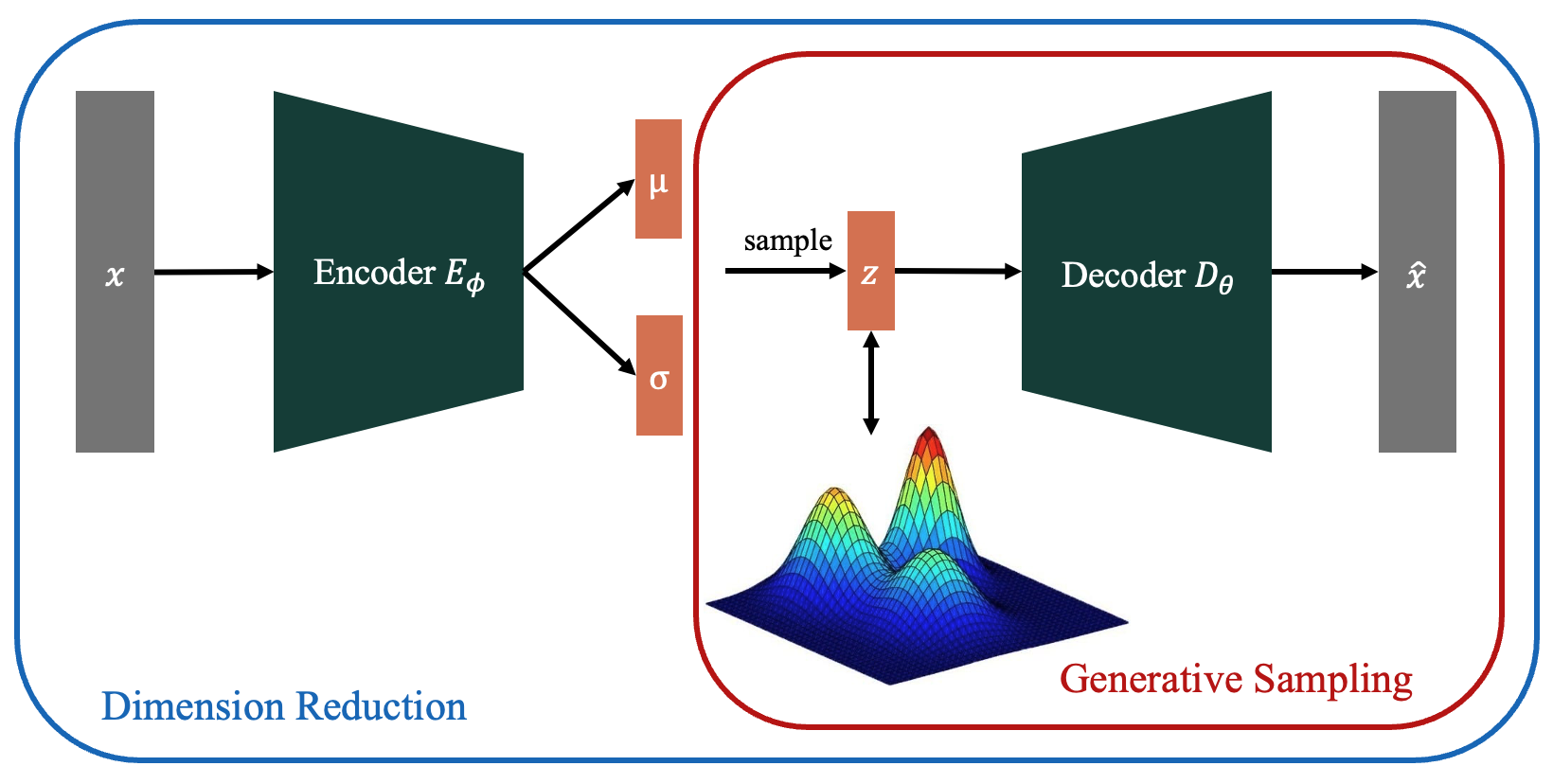}
    \caption{The GMVAE framework enables both dimension reduction and generative sampling.}
    \label{fig:model}
\end{figure}
The GMVAE pipeline \Cref{fig:model} integrates a standard Variational Autoencoder (VAE) \cite{kingma2014autoencoding} with a Gaussian Mixture Model (GMM) in the latent space.
Let $\text{Cat}(K, \boldsymbol{\pi})$ denote the categorical distribution, we assume:
\begin{align*}
    c & \sim \text{Cat}(K, \boldsymbol{\pi}), \\
    \mathbf{z} \mid c & \sim \mathcal{N}(\boldsymbol{\mu}_c, \boldsymbol{\sigma}^2_c \mathbf{I}), \\
    \mathbf{x} \mid \mathbf{z} & \sim \mathcal{N}(\tilde{\boldsymbol{\mu}}, \tilde{\boldsymbol{\sigma}}^2 \mathbf{I}),
\end{align*}
where \( K \) is the number of clusters, $c$ is the cluster label, and \( \boldsymbol{\mu}_c, \boldsymbol{\sigma}^2_c \mathbf{I} \) are the mean and covariance matrix of each cluster.

\subsection{Loss Function}
\label{subsec:loss_function}

The GMVAE training optimizes the Evidence Lower Bound (ELBO) \cite{jordan1999introduction} of the log-likelihood of the data, expressed as:
\begin{align*}
    \mathcal{L}_{\text{ELBO}}(\mathbf{x}) &= \mathbb{E}_{q(\mathbf{z} \mid \mathbf{x})} \left[ \log p(\mathbf{x} \mid \mathbf{z}) \right] 
    + \mathbb{E}_{q(\mathbf{z}, c \mid \mathbf{x})} \left[ \log p(\mathbf{z} \mid c) \right] \\
    & + \mathbb{E}_{q(c \mid \mathbf{x})} \left[ \log p(c) \right] - \mathbb{E}_{q(\mathbf{z} \mid \mathbf{x})} \left[ \log q^{(\mathbf{z})}(\mathbf{z} \mid \mathbf{x}) \right] \\
    & 
    - \mathbb{E}_{q(c \mid \mathbf{x})} \left[ \log q^{(c)}(c \mid \mathbf{x}) \right],
\end{align*}
where we adopt the mean-field approximation \cite{jordan1999introduction} \( q(\mathbf{\mathbf{z}}, c \mid \mathbf{x}) \approx q^{(\mathbf{z})}(\mathbf{z} \mid \mathbf{x}) q^{(c)}(c \mid \mathbf{x}) \) to simplify the optimization \cite{jiang2016variational}.

\subsection{Training Strategy}
\label{subsec:training_strategy}

We optimize two subgroups of the model parameters in an alternating block descent manner. The encoder-decoder network parameters and the GMM cluster priors $\boldsymbol{\pi}$ are updated via gradient descent, while the GMM cluster means $\boldsymbol{\mu}_c$ and variances $\boldsymbol{\sigma}^2_c$ are updated using the Expectation-Maximization (EM) algorithm \cite{dempster1977maximum}. 



\subsection{Evaluation Metric for Interpretability} \label{subsec:metric}

To achieve physical interpretability in latent representations, it is essential that continuous physical quantities, which characterize the states of the physical system, vary smoothly with respect to the latent coordinates.

Leveraging graph spectral theory \cite{chung1997spectral}, which analyzes smoothness of spatially-varying quantities through the eigenmodes of the graph Laplacian, we develop a quantitative metric to assess the interpretability of dimension reduction methods. A graph is constructed using a 
k-nearest neighbors (k-NN) approach based on two-dimensional PCA-transformed latent embeddings, capturing local geometry. The graph Laplacian’s eigenvectors represent modes of variation, with smaller eigenvalues corresponding to smoother modes. Physical quantities (e.g., pressure) are projected onto these eigenvectors, and the energy concentration in the top 
$\alpha$ fraction of smoothest modes is computed, reflecting how well the latent manifold preserves global physical structures. 

%% file: GMVAE-dimension-reduction/sections/5-experiments.tex
\section{Experiments and Discussions}

We simulated data from the classic 2D "flow past a cylinder" benchmark, generated with the FEniCS project \cite{Alnaes2015fenics}. Each data sample includes three channels: velocity in the x-direction (\( u \)), velocity in the y-direction (\( v \)), and pressure (\( p \)), capturing flow regimes characterized by Reynolds numbers (\( Re \)) uniformly distributed between 98 and 2000. The flow fields are taken at time $t=5.0s$ for each Reynolds number. 
Gaussian noise with a standard deviation at 15\% of the data magnitude was added. 

\begin{figure}[tbh!]
    \centering
    \begin{minipage}{0.31\linewidth}
        \centering
        {\scriptsize (a) Isomap}\\
        \includegraphics[width=\linewidth, trim=0 0 5cm 0, clip]{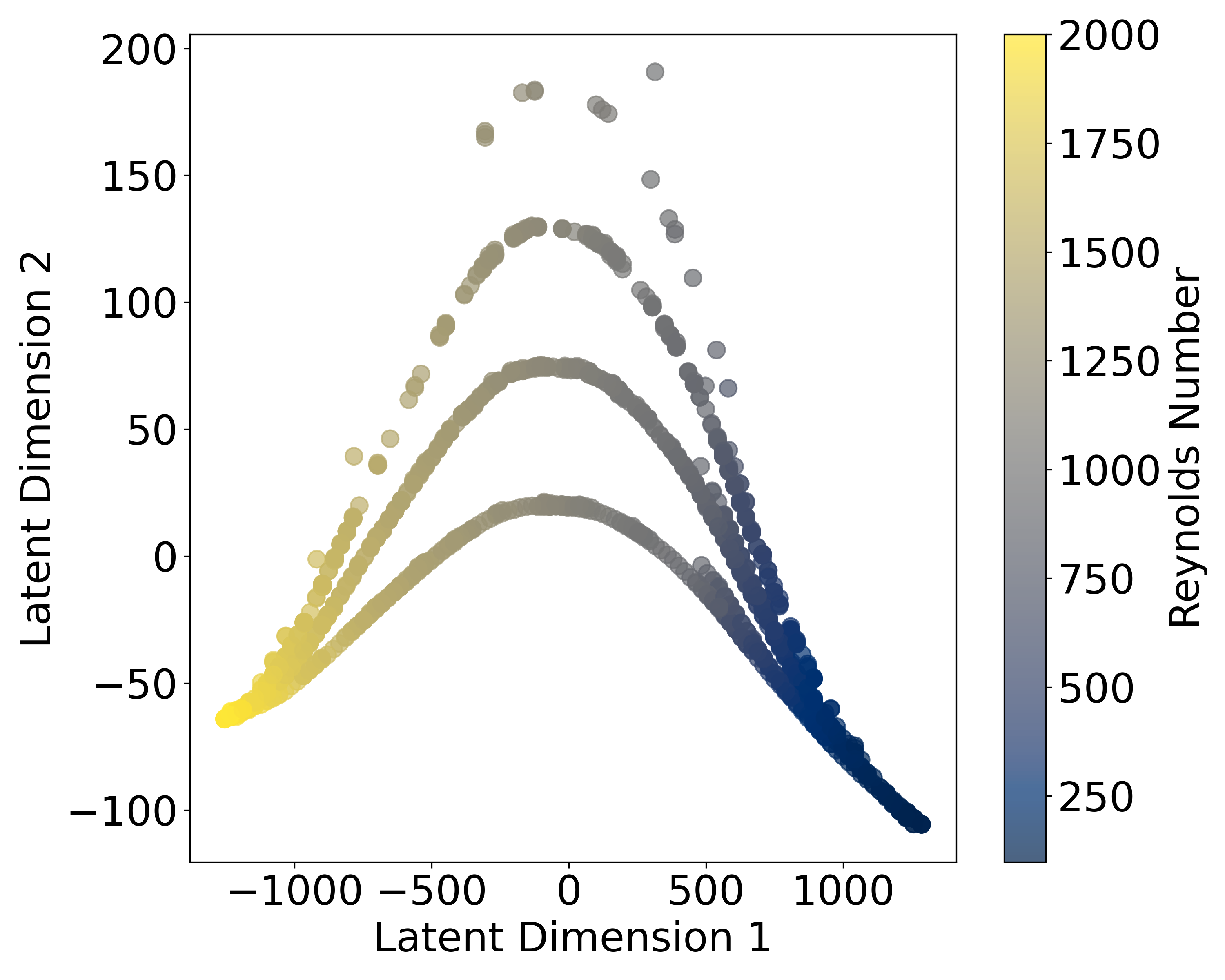}
    \end{minipage}
    \hfill
    \begin{minipage}{0.31\linewidth}
        \centering
        {\scriptsize (b) UMAP}\\
        \includegraphics[width=\linewidth, trim=0 0 5cm 0, clip]{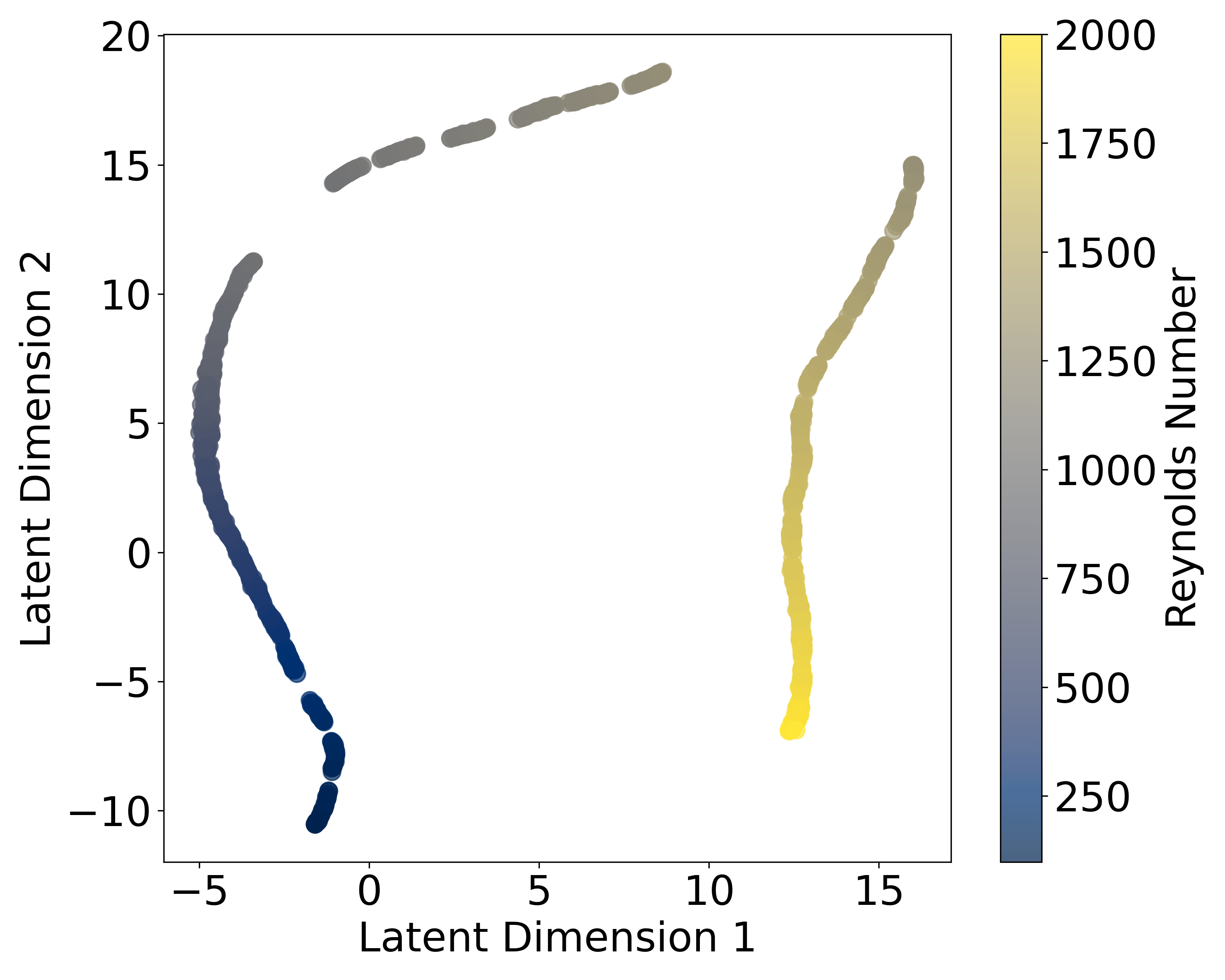}
    \end{minipage}
    \hfill
    \begin{minipage}{0.31\linewidth}
        \centering
        {\scriptsize (c) VAE}\\
        \includegraphics[width=\linewidth, trim=0 0 5cm 0, clip]{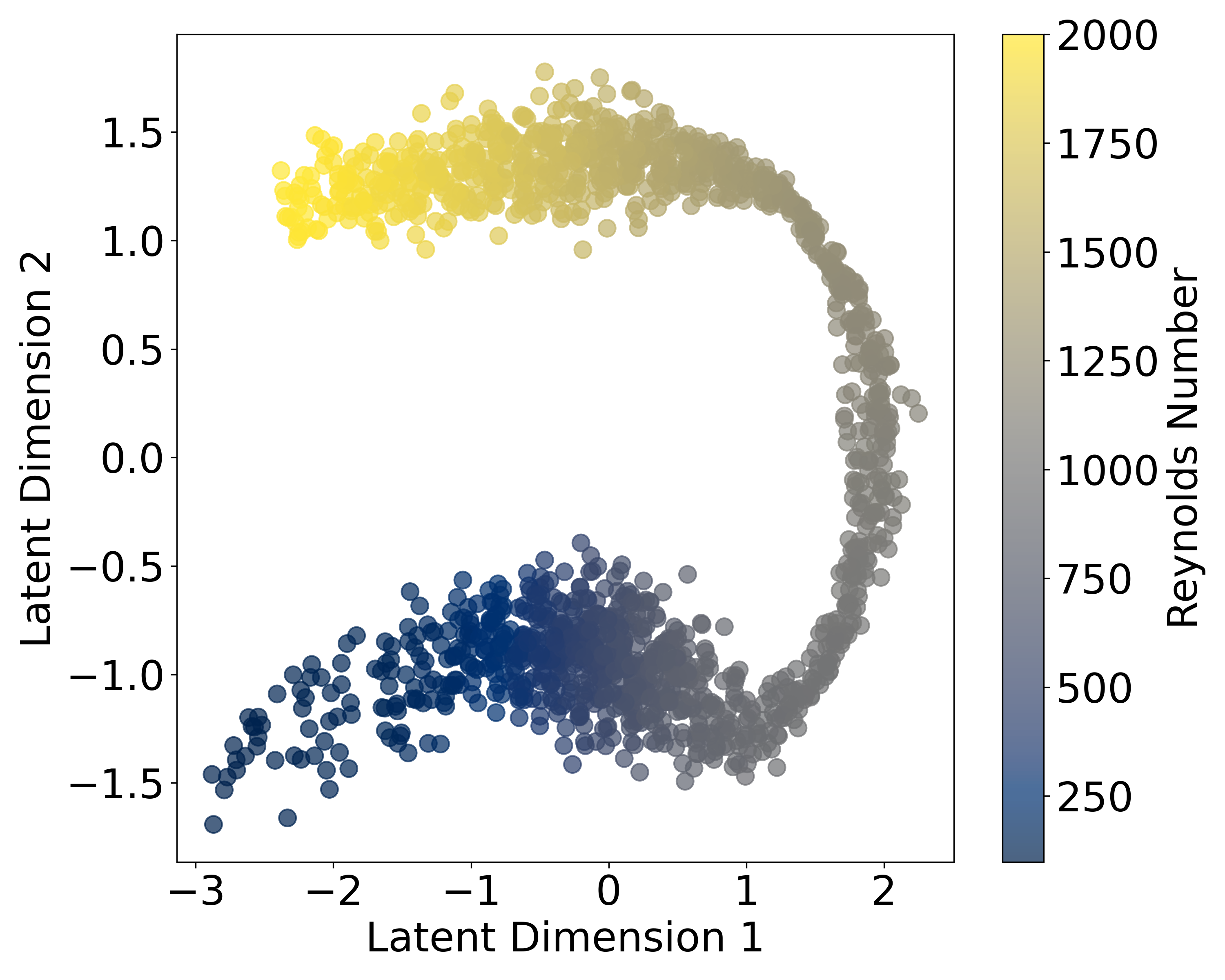}
    \end{minipage}
    
    \vspace{0.3cm}
    
    \begin{minipage}{0.37\linewidth}
        \centering
        {\scriptsize (d) GMVAE}\\
        \includegraphics[width=\linewidth]{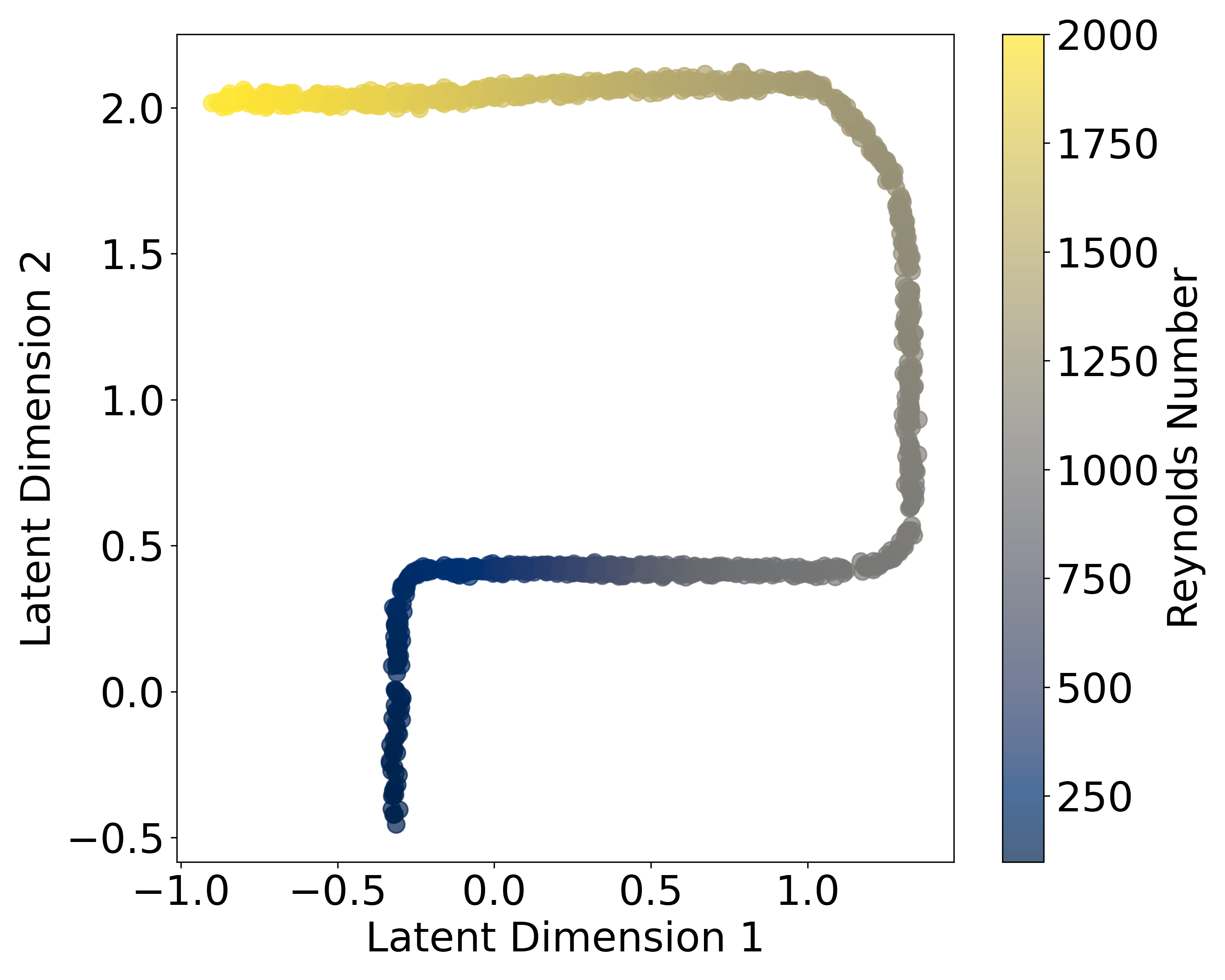}
    \end{minipage}
    \hfill
    \begin{minipage}{0.28\linewidth}
        \centering
        {\scriptsize (e) Interpretability} \vspace{0.4cm}\\ 
        \includegraphics[width=\linewidth, trim=0 0 0 0, clip]{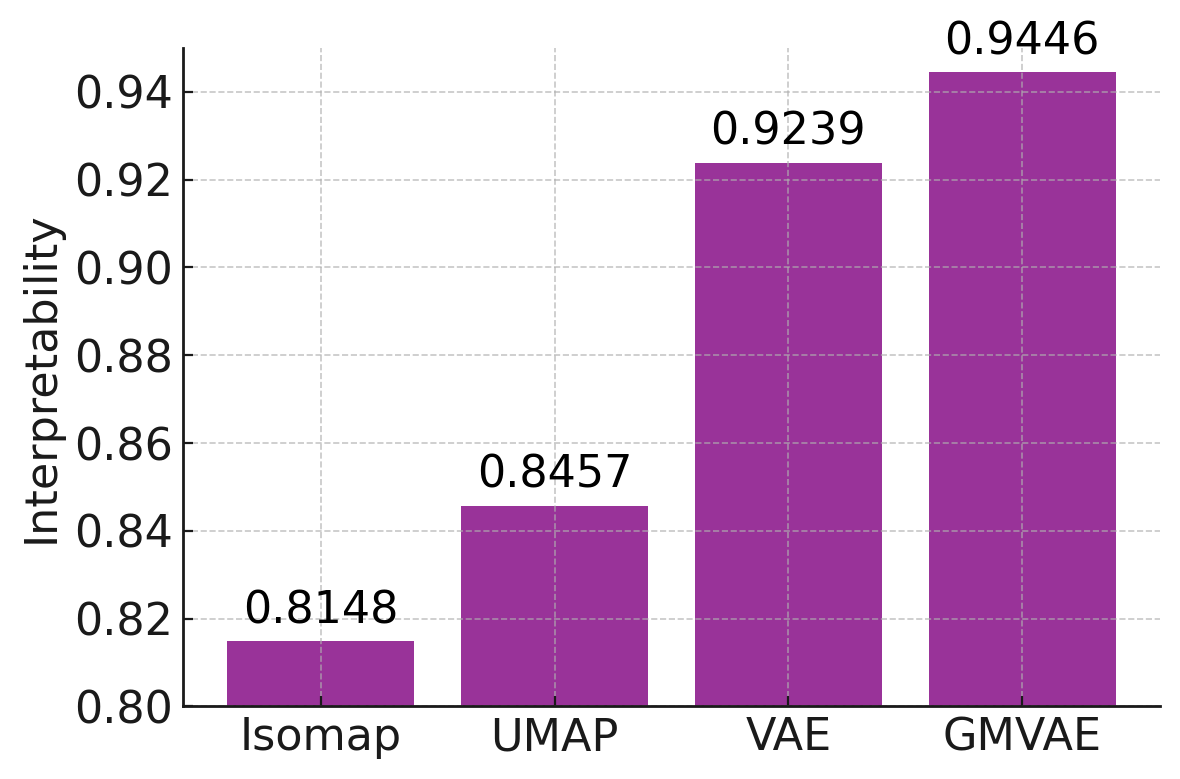}\vspace{0.4cm}
    \end{minipage}
    \hfill
    \begin{minipage}{0.32\linewidth}
        \centering
        {\scriptsize (f) GMVAE Cluster Labels}\\
        \includegraphics[width=\linewidth, trim=0 0.3cm 0 0, clip]{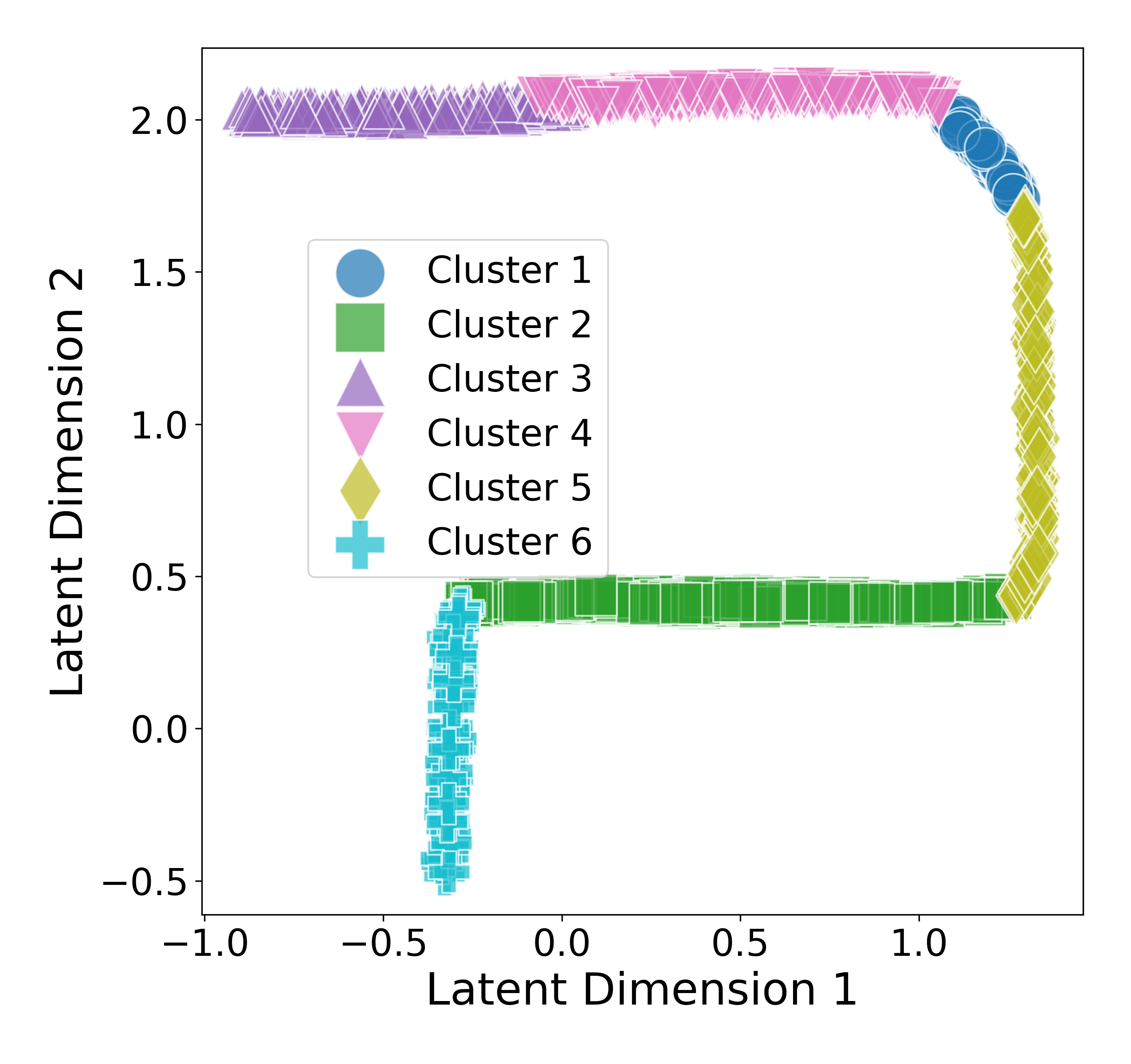}
    \end{minipage}
    
    \vspace{0.1cm}
    
    \begin{minipage}{\linewidth}
        \centering
        {\scriptsize (g) GMVAE Cluster Centroids}\\
        \includegraphics[width=1.0\linewidth, trim=0 0 0 1.1cm, clip]{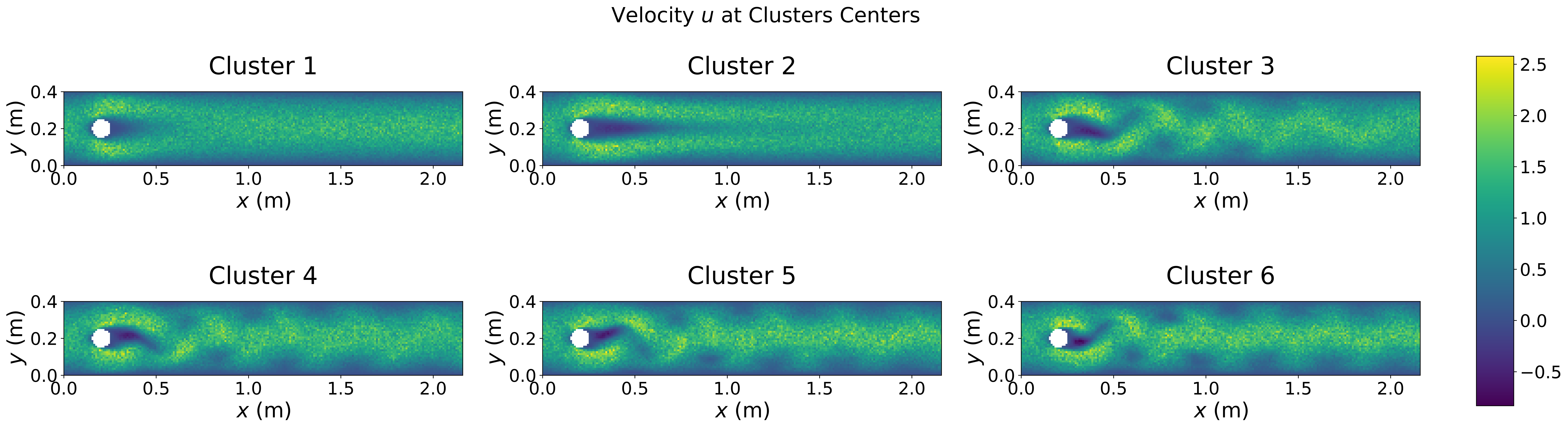}
    \end{minipage}
    
    \caption{GMVAE latent manifold of Navier-Stokes flow fields showing a continuous distribution of Reynolds numbers, with clusters corresponding to distinct physical states.}
    \label{fig:navier_stokes_latent_manifold}
\end{figure}

To establish a baseline, we applied the isometric feature mapping (Isomap)\cite{tenenbaum2000global}, 
Uniform
Manifold Approximation and Projection (UMAP)\cite{mcinnes2018umap}, and VAE to reduce the dataset to 2 latent dimensions (\Cref{fig:navier_stokes_latent_manifold}). The first two methods effectively preserve pixel-domain and latent space similarity but struggle to separate data from different Reynolds numbers, revealing their limitations in high-dimensional spaces. The baseline VAE, constrained by its Gaussian assumption for the latent variables, tends to cluster data globally, merging data points with distinct Reynolds numbers. 

We fit a GMVAE with a U-Net-inspired convolutional encoder-decoder architecture and a two-dimensional latent space on this dataset. As shown in \Cref{fig:navier_stokes_latent_manifold}, the GMVAE’s Gaussian mixture model assumption enables a more disentangled latent manifold compared to the baseline VAE.
The latent manifold's smooth distribution of Reynolds numbers suggests that the GMVAE effectively encodes the continuous nature of fluid flow regimes. Our quantitative interpretability metric from \cref{subsec:metric} (with $k=10$, $\alpha=5\%$) validates the observation. The GMM cluster centroids correspond to characteristic simulations at various Reynolds number levels. The GMVAE embeddings exhibit robustness to variations in the number of
clusters, provided the number of clusters is sufficiently large (e.g., larger than 3 for this dataset). 

Furthermore, the pipeline enables conditional generation of flow data. While fixing encoder, decoder, and GMM, we trained a separate Multilayer Perceptron (MLP) to map Reynolds numbers to the latent embeddings. Combining the MLP and the decoder created a generative pipeline that transforms a Reynolds number into velocity and pressure fields.

%% file: GMVAE-dimension-reduction/sections/6-conclusion.tex
\section{Conclusions}

The GMVAE framework effectively addresses dimension reduction, clustering, and generative sampling while preserving global physical similarities in the data, enabling meaningful and interpretable latent representations. It demonstrates robust generative capabilities and high clustering accuracy, making it valuable for analyzing complex systems such as combustion processes and fluid dynamics. 
Future work aims to extend GMVAE to multi-modal learning tasks by integrating diverse data sources and enabling cross-modal translation.